# Strong interlayer charge transfer due to exciton condensation in an electrically-isolated GaAs quantum well bilayer

[1,2,3] Joonho Jang*, [1]Heun Mo Yoo, [4]Loren N. Pfeiffer [4]Kenneth W. West, [4]K. W. Baldwin, [1]Raymond C. Ashoori*

1 Department of Physics, Massachusetts Institute of Technology, Cambridge, MA 02139, USA

2 Department of Physics and Astronomy, and Institute of Applied Physics, Seoul National University, Seoul 08826, Korea

3 Center for Correlated Electron Systems, Institute for Basic Science, Seoul 08826, Korea

4 Department of Electrical Engineering, Princeton University, Princeton, NJ 08544, USA

*correspondence to: joonho.jang@snu.ac.kr, ashoori@mit.edu

## Abstract

We introduce a design of electrically isolated "floating" bilayer GaAs quantum wells (QW) in which application of a large gating voltage controllably and highly reproducibly induces charges that remain trapped in the bilayer after removal of the gating voltage. At smaller gate voltages, the bilayer is fully electrically isolated from external electrodes by thick insulating barriers. This design permits full control of the total and differential densities of two coupled 2D electron systems. The floating bilayer design provides a unique approach for studying systems inaccessible by simple transport measurements. It also provides the ability to measure the charge transfer between the layers, even when the in-plane resistivities of the 2D systems diverge. We measure the capacitance and inter-layer tunneling spectra of the QW bilayer with independent control of the top and bottom layer electron densities. Our measurements display strongly enhanced inter-layer tunneling current at $v_T = 1$, a signature of exciton condensation of a strongly interlayer-correlated bilayer system. With fully tunable densities of individual layers, the floating bilayer QW system provides a versatile platform to access previously unavailable information on the quantum phases in electron bilayer systems.

Weakly coupled quantum Hall (QH) bilayer systems display a variety of exotic phases as the filling factor of the two layers vary [1–3]. Under quantizing magnetic fields, decreasing the separation distance of two layers to become comparable to the mean interparticle spacing can drive the double layer system into phases not possible in single layers [4–6]. When the individual layers are in strongly intra-layer correlated states such as fractional quantum Hall phases, the additional interlayer coupling can play a non-trivial role that modifies the properties of the phases or even result in totally new ground states of the coupled 2D system. There are many theoretical proposals for non-trivial QH phases in such vertically couple systems, including Pfaffian, Halperin, and exciton condensed phases [7–10].

The exciton condensate is an interlayer-coherent phase that shows dramatic macroscopic quantum phenomena, such as Josephson-like interlayer tunneling, quasiparticle tunneling and perfect Coulomb drag, as observed in GaAs quantum wells (QW) bilayers and graphene bilayers [1,11–15]. The interlayer interaction has an energy scale of $E_c = 1/4\pi\epsilon d$ and competes with an intralayer Coulomb interaction of energy $E_b = 1/4\pi\epsilon l_B$, where $d$ is the interlayer distance between the two layers and $l_B$ is the magnetic length. As $d/l_B$ decreases, the interlayer interaction starts to dominate over the intralayer interaction, favoring the formation of the interlayer correlated phases. The existence of the condensate phase depends critically on the ratio $d/l_B$, and experimenters have reported condensate phases at total filling factor $\nu_{tot} = 1$ ($\nu = 1/2$ in each layer) in samples with $d/l_B$ smaller than 1.8, in the limit of weak interlayer tunnel coupling [4,16,17]. In this picture, an interlayer bias voltage plays a role of pseudospin Zeeman energy for bilayer charge polarization.

In most previous experiments with bilayer QWs, charge carriers were induced by doping from top and bottom of the layers and one eventually needed to make ohmic contacts for transport measurements [2,18]. Specifically, the characteristic interlayer Josephson tunneling effect has been observed only with separate ohmic contacts to individual layers. With such a sample design, in-plane and contact resistances make high frequency measurements difficult, and it becomes impossible to make tunneling measurements when the longitudinal conductivity drops to zero. We instead use a contactless measurement scheme that functions independent of the conductivity of the layers. We electrically isolate the system of interest, a bilayer QW, in-between top and bottom insulating barriers and introduce charge carriers via a controlled leakage over the barriers by applying large bias voltages, and we find that the charge carriers persist in the double well system for long periods of time (hours or days) after the removal of the large bias. In this way, we are able to control the individual layer densities while monitoring the interlayer charge transfer without electrical ohmic contacts by detecting the displacement fields generated by interlayer-moving charges.

In this article, we present our contactless and electrically-isolated design approach of a bilayer QW with fully tunable top and bottom layer densities. Our results show that the exciton condensation occurs in the isolated bilayer device. The new design, along with capacitive pulsed tunneling spectroscopy [19] permits contactless measurement of the tunneling current along with direct wide-bandwidth probing of the sample needed for study of quantum coherence.

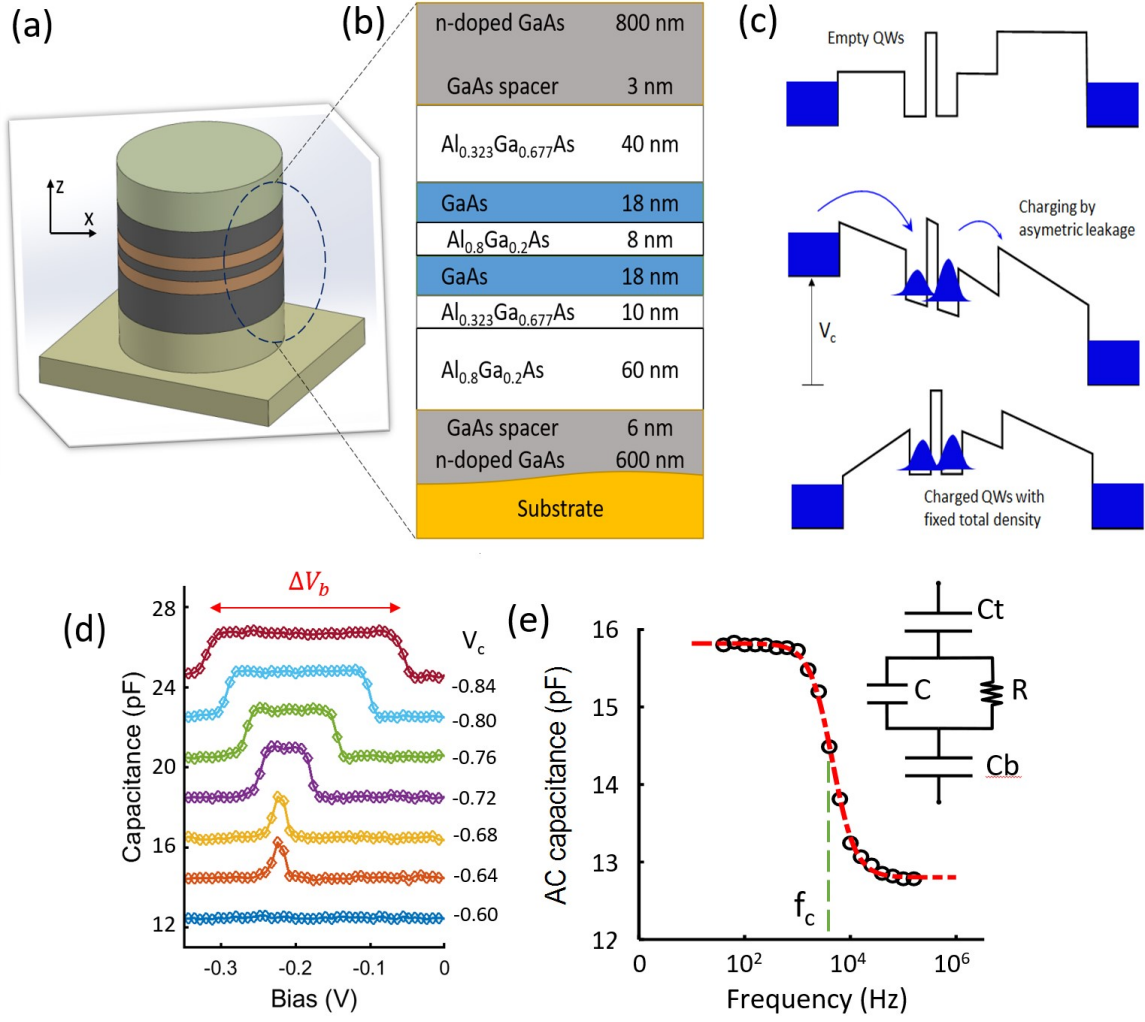

**Figure 1| Sample design and means for controlling total and differential density.** (a) Schematic of a cylindrically etched device. (b) The wafer growth profile used in this work. The GaAs QWs of 18 nm width are tunneling-coupled to each other and form a QW bilayer. (c) Sequence of charging the bilayer system. An initially empty bilayer is charged by first inducing leakage current over the structure, and capturing the charges in the bilayer. (d) Capacitance measured at 13 Hz as a function of $V_b$ at various values of $V_c$ used in the charging sequence of (c). Evidenced by the range of the high capacitance (red arrow), the amount of the total charge in the system grows with the increasing magnitude of $V_c$. Curves are shifted for clarity. (e) A frequency dependent capacitance measurement shows a roll-off behavior with the frequency ($f_c = 1/RC$) proportional to the strength of interlayer charge transfer conductance ($S_{in} = 1/R$). The red dotted line is a fit to the frequency response function of the effective circuit in the inset.

In **Fig. 1(a-b)** shown are the device schematic and the profile of the wafer (wafer id# pf-3.21.16) used in measurements. The DQW are composed of two 18 nm GaAs QWs separated by 6 nm $_{0.8}$AlGa$_{0.2}$As barrier. Thick barriers of Al$_{0.323}$Ga$_{0.677}$As and Al$_{0.8}$Ga$_{0.2}$As isolate the QWs from the external top and bottom electrodes (n-dope GaAs). The wafer is etched by photo-lithography to define a cylindrical mesa of 150 $\mu m$-diameter and 1.5 $\mu m$-height vertical device. The top and bottom electrodes are contacted to the external measurement setups. A capacitance bridge and HEMT amplifiers are used to enhance the sensitivity of the measurements in cryogenics environment [19,20].

The conduction band edge of the barriers, which determines the potential profile for induced electrons, are controlled by the Al content of $Al_xGa_{1-x}As$. The barriers have an asymmetric profile in top and bottom sides of the QWs by design so that electrons can accumulate in the bilayer when a large voltage bias $V_c$ is applied. The barriers are sufficiently thick to hold the induced charges in the bilayer system over several weeks once the large voltage bias is removed. In **Fig. 1(c)**, a sequence of charging the bilayer system is shown. An initially empty QWs are charged by an external voltage bias ($V_c$) that is sufficiently large to develop a leak current through the device. When the bias is removed, at cryogenic temperatures, we observe the charges in QW bilayers to remain sufficiently long so that we can perform complete measurements at a fixed total density to within our experimental resolution. We find that we can consistently and precisely set the amount of this remnant charge by controlling the value of $V_c$ that we apply. Then, a smaller bias voltage $V_b$ is used to control the density imbalance between the top and bottom QWs. Thus, with combinations of $V_c$ and $V_b$, the individual densities of the two QWs are independently tunable. Note that the range of bias voltage ($V_b$) is maintained much smaller than that of $V_c$ to prevent charges in the electrode or in the DQW from undergoing Fowler-Nordheim tunneling through one of the thick barriers, in which case the total charges in the bilayer can inadvertently change.

We designed our device structure to produce high mobility 2D electron systems; there is no doping present near the QWs and, after charging the double-well system with electrons. Upon removing $V_c$, the bilayer system remains electrically well isolated without any ohmic contacts from the external electrodes. To study charge transfer between the two layers embedded in the sample without contacts, we employ capacitance measurements and pulsed tunneling spectroscopy. For capacitance measurements, in addition to the DC bias voltages, we apply sinusoidal voltages with a fixed frequency and measure capacitive response between the electrodes using a home-made cryogenic capacitance bridge and HEMT amplifiers to study interlayer charge transfer. For tunneling spectroscopy, we use a contactless pulsed tunneling method, applying a sudden voltage step from external electrodes and remotely sense the electric field emanating from charges moving in-between layers via the HEMT amplifiers to measure interlayer tunneling spectra [19,20].

Precise determination of the total and differential density requires accurate measurements of capacitance of the device. We define total density $n_{tot} = n_1 + n_2$ and differential density $n_d = n_1 - n_2$, where $n_1$ and $n_2$ are the densities of top QW and bottom QWs, respectively. Once the total density of electrons in the bilayer system is fixed by the charging method described in **Fig. 1(c)**, we can vary $V_b$ to partition the remnant electron density between the wells, up to full depletion of either one of the QW layers. In **Fig. 1(d),** we plot capacitance data measured at quasi DC limit of $f = 13\ Hz$ at various charging voltages $V_c$. The region of high capacitance values (see the red arrow in the figure) indicates the existence of interlayer charge transfer when both layers have nonzero densities and their Fermi levels match, thus making the bilayer charge-polarizable. As the total density increases with a larger magnitude of $V_c$, the voltage range of the high capacitance ($\Delta V_b$) becomes wider because more total charges in the system require larger positive or negative $V_b$ to deplete either top or bottom QW. The data are consistent with independent control of the densities of two layers through changing $V_b$ and $V_c$.

In the region of high capacitance, if we increase the measurement frequency, the measured value for capacitance shows a roll-off behavior with the characteristic frequency, $f_c$, as in **Fig. 1(e)**. This frequency dependent capacitance (or AC capacitance) data yields information about the rate of charge movement between QW layers. In the zero-frequency limit, the capacitance measures thermodynamic charge-polarizability of the bilayer but, as the frequency increases, the measured capacitance falls because the interlayer charge transfer cannot keep up with the AC excitation due to finite interlayer tunneling conductivity. If we parameterize the interlayer charge transfer rate using a resistance $R$ of an effective circuit diagram representing the device in the inset, $f_c$ is given by $1/\sqrt{RC}$, where $C$ is the geometric capacitance of the bilayer.

In **Fig.2**, the AC capacitance measured as a function of $V_b$ and $V_c$ displays a clear signature of density control in an applied magnetic field. With increasing perpendicular magnetic fields, Landau levels form. Because filled Landau levels cannot charge due to energy gaps, we observe stripes of a suppressed capacitance when the Landau levels are full. This provides straight-forward assignment of filling factors ($\nu = n\,h/eB$) to top and bottom QWs for the incompressible stripes, and they are denoted in the figure. At magnetic fields at and above 6T, we observe that visible incompressible stripes and areas appear at fractional filling factors such as 1/3, demonstrating that the electrically-floating contactless design offers a versatile platform for studying the coupled systems of two single-layer QH phases.

We focus in the rest of this paper on the exciton condensed phase [1,11,12]. When we increase the frequency of the capacitance measurement to 100 kHz, charge transfer is suppressed except for the regions where the interlayer charge transfer is comparably faster. As discussed above, if the characteristic frequency $f_c$ is much smaller than the measurement frequency $f$, the AC capacitance signal would become strongly suppressed. First of all, for $B = 0\,T$, the thin line of high interlayer charge transfer stands out (**Fig. 2 (b)**) and this line corresponds to the equal densities of the two QWs; here, the conduction band edge and the energy dispersion of the two layers match, so that momentum conservation for interlayer tunneling is well satisfied, leading to fast interlayer tunneling [19,20]. However, upon application of magnetic fields, Landau quantization modifies the condition for the momentum conservation and leads to the appearance of multiple patches of high capacitance. It is well known that Coulomb interaction between electrons creates a Coulomb gap at the Fermi level [21,22] near non integer filling factors, leads to an exchange gap around odd integer filling factors, suppressing tunneling near zero bias (i.e. Fermi level) [18]. Consequently, we attribute the small patches of fast charge transfer in 100 kHz data as a consequence of the Coulomb gap and the exchange gap disappearing near even integer filling factors.

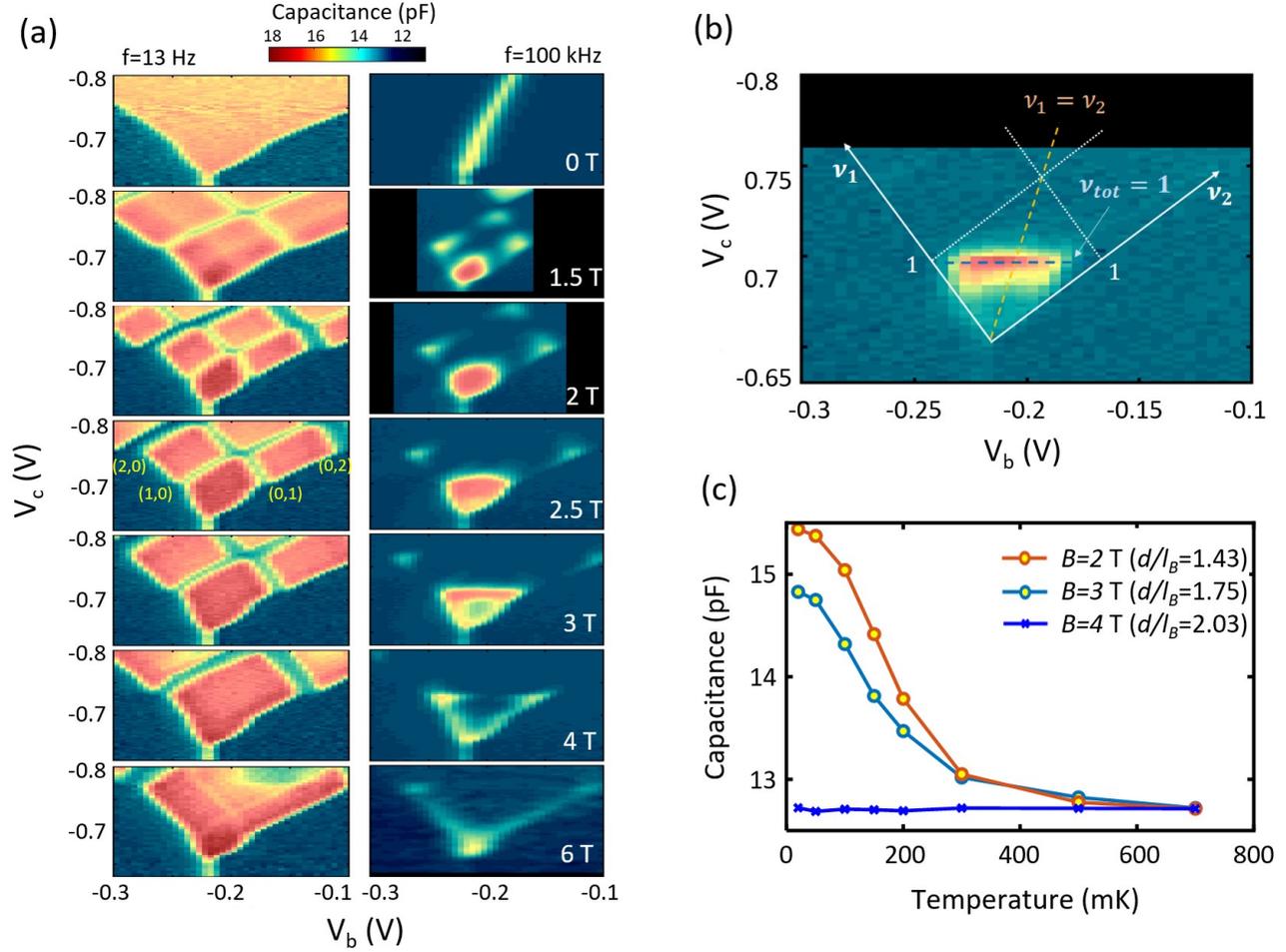

**Figure 2| Magnetic field and frequency dependence of AC capacitance measurements. (a)** At base temperature of 20 mK, AC capacitances are plotted as a function of charging voltage ($V_c$) and bias voltage ($V_b$), at various magnetic fields and two different frequencies. The interlayer charge transfer rate is fast at certain filling factors due to the increased correlation of the top and bottom layers. **(b)** AC capacitance measured at f=1 MHz and B=2.5 T. Total and individual filling factors ($\nu_{tot}, \nu_t, \nu_b$) are denoted. At the total filling factor of 1 (dotted horizontal line), the very high interlayer conductivity signals the formation of the exciton condensation. **(c)** AC capacitance measured at $\nu_t=1/2$, $\nu_b=1/2$ with f=100 kHz strongly increases at low temperature when $d/l_B < 1.8$. This temperature dependence suggests enhanced interlayer charge transfer as a result of the exciton condensation.

However, noticeably at $B = 2.5\ T$ and $3\ T$, the signal becomes very strong along the line that corresponds to $\nu_T = 1$, suggestive of a strongly correlated top and bottom QH phase. The signal disappears for higher magnetic fields at 4 and 6 $T$. In **Fig. 2 (b)**, we observe that, for $B = 2.5\ T$, the AC capacitance (a measure of interlayer charge transfer rate) at $\nu_T = 1$ remains high even at $f = 1\ MHz$, the highest frequency in our capacitance measurements. We attribute this ultra-fast interlayer charge transfer to very high zero-bias tunneling conductivity as a consequence of the formation of an exciton condensed phase in the bilayer QW system. This fast interlayer charge transfer suggests the presence of Josephson tunneling of the emergent interlayer coherent phase [4]. Because the AC voltage amplitude applied between the QWs for the capacitance measurements is only 100 $\mu V$, the signals at a very high frequency likely arise from the interlayer zero-bias conductivity peak, such as the nearly dissipationless Josephson current of a bilayer exciton condensate [16]. Furthermore, as shown in **Fig. 2 (c)**, signals have a strong temperature dependence, and the feature only appears below 300 mK with $d/l_B \leq 1.8$, consistent with previous reports [16,17].

To confirm the picture, we performed the time-resolved tunneling measurements of the interlayer charge transfer in **Fig 3**. We used square-shaped excitations on the external electrodes and measured interlayer charge transfer over a tunneling barrier in a time domain, following the techniques described in previous studies [19,20]. When the spectrum is measured at $B = 2.5\ T$ and $T = 20\ mK$, the tunneling conductivity is observed to be strongly peaked near zero tunneling voltage, and its width is as narrow as $100\ \mu V$. However, when temperature is increased to $700\ mK$, the peak becomes significantly suppressed. Increasing the magnetic field rather than the temperature also affects the signal. The data at $4\ T$ and $20\ mK$ show that the strong peak disappears, while the typical spectral features of the aforementioned Coulomb gap near zero tunneling bias remain visible (**Fig. 3 (b)**). Thus, the tunneling spectra gives fully consistent pictures with the AC capacitance data of the Josephson tunneling peak.

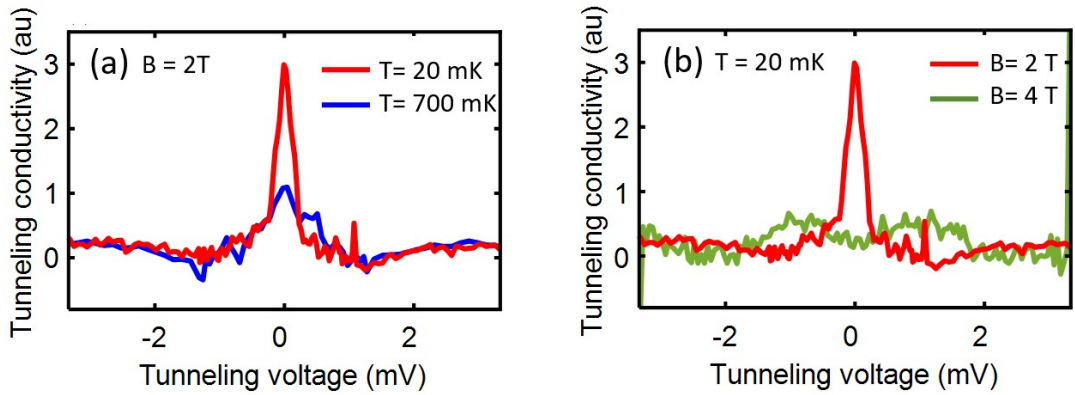

**Figure 3 | Tunneling spectra of an exciton condensed phase.** (a) Time-resolved tunneling spectra measured at B = 2T with T = 20 mK (red) and 700 mK (blue). (b) Spectra measured at T = 20 mK with B = 2T (red) and 4T (green). Note that the red curve is the same in (a) and (b) for better comparison. The strong peak near zero tunneling bias suggests Josephson effect of an interlayer coherent exciton phase.

One of the main advantages of this electrically isolated "floating" bilayer design becomes more apparent when considering the range of differential densities over which we can observe the enhanced interlayer charge transfer of an exciton phase. Unlike previous transport measurements that showed evidence of this exciton condensation and its Josephson-like tunneling effects [16,17], our samples are designed to induce and measure vertical charge transfer independent of the lateral conductivities of individual 2D systems. This capability can be quite crucial when tuning densities of 2D systems in regimes of highly incompressible QH phases [24]. We can perform experiments with large differential density of the QW bilayer; i.e. the densities (and thus filling factors) of two QWs can be tuned in a wide range independently, even to incompressible low conductive regions, while maintaining the accuracy of the measurements. Perhaps more importantly, these measurements can be done without direct external contacts to any of these layers, eliminating any chance that the measurements are affected by extrinsic effects, so that a delicate quantum phase, such as the exciton condensation phase, is best preserved to provide evidence as a bulk phenomenon and a thermodynamic ground state. The strong signal at high frequency, such as shown in **Fig.2 (b)**, in a wide range of differential density along the dotted line of $\nu_T = 1$ indicates the robust formation of the condensate in our sample. More detailed study of the exciton condensates in large density imbalance and their tunneling spectra will be presented elsewhere [25].

Strongly interacting 2D bilayers provide a versatile platform to engineer rich physics as parameters for interlayer and intralayer interactions are tuned. The sample design and technique presented in this paper allow control of the parameters in a wide range, and lead to the realization of the exciton condensation, demonstrating as a promising direction to control and detect the quantum phases of bilayer in electrically isolated devices. Considering recent progress in new material systems such as graphene and transition metal dichalcogenides towards fabricating high quality bilayers, our unique approach presented here will be found to be useful [26,27].

## ACKNOWLEDGEMENT

The work at MIT was supported by the Basic Energy Sciences Program of the Office of Science of the US Department of Energy through contract no. FG02-08ER46514 and by the Gordon and Betty Moore Foundation through grant GBMF2931. The work at Princeton was by the Gordon and Betty Moore Foundation's EPiQS Initiative, Grant GBMF9615 to L. N. Pfeiffer, and by the National Science Foundation MRSEC grant DMR 1420541. The work at SNU was supported by the Creative-Pioneering Researchers Program of Seoul National University, the POSCO Science Fellowship of POSCO TJ Park Foundation and the National Research Foundation (NRF) of Korea (Grant no. 2020R1A5A1016518 and 2019R1C1C1006520).

**Data availability**

The data that support the findings of this study are available from the corresponding author upon reasonable request.

**Conflict of Interest**

Authors declare no conflict of interest.